\documentclass[showpacs,10pt,twocolumn,prb]{revtex4-1}
\usepackage{amsmath}
\usepackage{amssymb}
\usepackage{graphics}
\usepackage{epsfig}
\usepackage{CJK}
\usepackage{color}

\begin{document}


\title{Massive fermions with low mobility in antiferromagnet orthorhombic CuMnAs single crystals}
\author{Xiao Zhang,$^{1}$ Shanshan Sun,$^{2}$ and Hechang Lei$^{2,}$}
\email{hlei@ruc.edu.cn}
\affiliation{$^{1}$State Key Laboratory of Information Photonics and Optical Communications $\&$ School of Science, Beijing University of Posts and Telecommunications, Beijing 100876, China
\\$^{2}$Department of Physics and Beijing Key Laboratory of Opto-electronic Functional Materials $\&$ Micro-nano Devices, Renmin University of China, Beijing 100872, China}

\date{\today}

\begin{abstract}

We report the physical properties of orthorhombic $o$-CuMnAs single crystal, which is predicted to be a topological Dirac semimetal with magnetic ground state and inversion symmetry broken. $o$-CuMnAs exhibits an antiferromagnetic transition with $T_{N}\sim$ 312 K. Further characterizations of magnetic properties suggest that the AFM order may be canted with the spin orientation in the $bc$ plane. Small isotropic MR and linearly field-dependent Hall resistivity with positive slope indicate that single hole-type carries with high density and low mobility dominate the transport properties of $o$-CuMnAs. Furthermore, the result of low-temperature heat capacity shows that the effective mass of carriers is much larger than those in typical topological semimetals. These results imply that the carriers in $o$-CuMnAs exhibit remarkably different features from those of Dirac fermions predicted in theory.

\end{abstract}


\maketitle


\section{Introduction}

Topological semimetals (TSMs), as a new kind of topological materials besides topological insulators (TIs)\cite{Hasan,QiXL}, topological superconductors (TSCs)\cite{QiXL}, and topological crystalline insulators (TCIs)\cite{FuL}, have attracted tremendous attentions in recent. The TSMs are characterized by non-trivial bulk band crossings (nodal points) with linear dispersion in the momentum space\cite{WanX,XuG,Burkov,Burkov2}. If doubly degenerate bands cross at a fourfold degenerate crossing (Dirac point) protected by the crystal symmetry, it will form a Dirac semimetal (DSM)\cite{YoungSM,WangZ,WangZ2}. The electronic structure of a three-dimensional (3D) DSM is a 3D analogue of graphene with two-dimensional Dirac points\cite{Neto}. Such kind of 3D DSMs has been theoretically predicted and experimentally confirmed in several materials, such as Na$_{3}$Bi, Cd$_{3}$As$_{2}$, and Pt(Te/Se)$_{2}$ etc\cite{WangZ,WangZ2,LiuZK,LiuZK2,Neupane,Borisenko,XuSY,HuangH,YanM}. On the other hand, when the doubly degenerate bands are lifted by breaking spatial inversion symmetry ($P$) or time reversal symmetry ($T$), a Dirac point splits to a pair of Weyl point with opposite chirality due to singly-degenerate band crossings and a DSM will evolve into a Weyl semimetal (WSM)\cite{WanX,Burkov,YoungSM,WangZ,WangZ2,WengH,HuangSM}. The WSMs with inversion symmetry broken have been shown in (Ta/Nb)(As/P)\cite{WengH,HuangSM,XuSY2,LvBQ,YangLX,XuSY3,XuN}, and (W/Mo)Te$_{2}$ etc\cite{Soluyanov,WuY,WangC,DengK}. The WSMs with spontaneous time reversal symmetry broken have also been reported in YbMnBi$_{2}$\cite{Borisenko2}, and predicted in magnetic HgCr$_{2}$Se$_{4}$ and Heusler alloys etc\cite{ChangG,WangZ3,Kubler}. The existence of Weyl points near the Fermi energy level ($E_{F}$) will lead to interesting spectroscopic and transport phenomena, such as Fermi arcs\cite{WanX,XuSY,LvBQ,XuSY2,DengK}, ultrahigh carrier mobility\cite{LiangT,Shekhar}, and chiral anomaly\cite{HuangXC,XiongJ,Hirschberger}.

Interestingly, recent theoretical studies predict that even when the $P$ and $T$ symmetries are broken the combined $PT$ symmetry can be preserved in the antiferromagnetic (AFM) orthorhombic CuMn(As/P) ($o$-CuMn(As/P))\cite{TangP,Smejkal}. When the magnetic moments of Mn are aligned along the $z$ axis and screw rotation symmetry $S_{2z}$ survives, the Dirac points can still exist with considering spin-orbital coupling (SOC). Furthermore, when the magnetic moments of Mn are along other directions, such as (111) and (101), the $S_{2z}$ symmetry will be broken and the Dirac points will be gapped. In that case, $o$-CuMnAs will become an AFM semiconductor\cite{TangP,Smejkal}. Thus, $o$-CuMnAs provides a good platform to investigate the interplay between Dirac fermions and AFM state as well as the topological metal-insulator transition (MIT) driven by the N\'{e}el vector reorientation\cite{TangP,Smejkal}. In contrast to theoretical studies, the experimental results of $o$-CuMnAs is relatively rare. A study on $o$-CuMnAs powder in the literature only indicates that there is an AFM transition above room temperature\cite{Maca}. Because of a lack of single crystals, whether this material could exhibit the behaviors of Dirac fermions is still unknown. In this work, we investigate the physical properties of $o$-CuMnAs single crystals. $o$-CuMnAs enters a canted long-range AFM ordering state at $T_{N}\sim$ 312 K and the spin orientation is in the $bc$ plane. Moreover, it shows a metallic behavior and single hole-type carriers dominate the transport properties. Surprisingly, $o$-CuMnAs exhibits relatively low carrier mobility, large effective mass and small magnetoresistance (MR), distinctly different from typical TSMs.

\section{Experimental}

$o$-CuMnAs single crystals were grown from Bi flux. The raw materials Cu shot (purity 99.9 \%), Mn pieces (purity 99.9 \%), As (purity 99.995 \%) and Bi lumps (purity 99.99 \%) were mixed together in a molar ratio of Cu : Mn : As : Bi = 1 : 1 : 1 : 8 and placed in an alumina crucible. The crucible was sealed in the quartz ampoule under partial argon atmosphere. The sealed quartz ampoule was heated up to 1173 K and kept there for 12 h to ensure the homogeneity of melt. Then, the ampoule was cooled down slowly to 723 K with 4 K/h and at this temperature, the flux was decanted with a centrifuge. Air-stable $o$-CuMnAs single crystals with typical size 3$\times$1$\times$0.1 mm$^{3}$ can be obtained. X-ray diffraction (XRD) pattern of an $o$-CuMnAs single crystal was measured using a Bruker D8 X-ray machine with Cu $K_{\alpha}$ radiation ($\lambda=$ 0.15418 nm) at room temperature. The lattice parameters were determined by single crystal XRD. The data were collected using the Bruker APEX2 software package \cite{APEX2} on a Bruker SMART APEX II single-crystal X-ray diffractometer with graphite-monochromated Mo $K_{\alpha}$ radiation ($\lambda=$ 0.071073 nm) at room temperature. The elemental analysis was performed using the energy-dispersive x-ray spectroscopy (EDX) analysis. Magnetization measurement was carried out in Quantum Design MPMS3. Electrical transport and heat capacity measurements were carried out in Quantum Design PPMS. Both longitudinal and Hall electrical resistivity were measured using a standard four-probe method on rectangular shape single crystals with current flowing in the $bc$ plane. In order to effectively avoid the longitudinal resistivity contribution due to voltage probe misalignment, the Hall resistivity was measured by sweeping the field from -9 T to 9 T at various temperatures, and the total Hall resistivity was determined by $\rho_{xy}(\mu_{0}H)=[\rho(+\mu_{0}H)-\rho(-\mu_{0}H)]/2$, where $\rho(\pm\mu_{0}H)$ is the transverse resistivity under a positive or negative magnetic field.

\section{Results and discussion}

\begin{figure}[tbp]
\centerline{\includegraphics[scale=0.35]{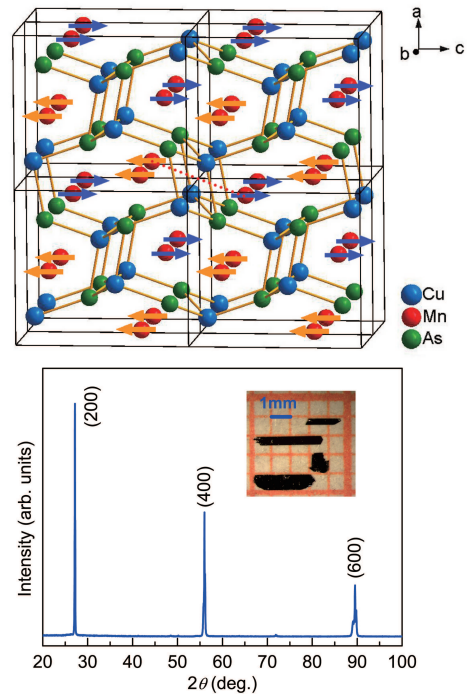}} \vspace*{-0.3cm}
\caption{(a) Crystal structure of $o$-CuMnAs. The big blue, medium red and small green ball represents Cu, Mn, and As atoms, respectively. The arrows locating at Mn atoms exhibit the spin directions predicted in theory. The dotted line connecting two Mn atoms demonstrates the preservation of $PT$ symmetry. (b) XRD pattern of an $o$-CuMnAs single crystal. Inset: photo of typical $o$-CuMnAs single crystals. The length of one grid in the photo is 1 mm.}
\end{figure}

$o$-CuMnAs has a TiNiSi-type structure with the non-symmorphic orthorhombic space group $Pnma$ (No. 62)\cite{Mundelein,Maca}. As shown in Fig. 1(a), in $o$-CuMnAs, Cu and As atoms form a three-dimensional four-connected anionic network. The network is composed of strongly corrugated two-dimensional sheets of edge-sharing six-membered rings. The Mn atoms fill the large channels left by this network along the $b$ axis. According to the results of theoretical calculation without considering SOC\cite{TangP}, the ground state of $o$-CuMnAs is antiferromagnetic and the magnetic configuration with the lowest energy is shown in Fig. 1(a). The spin orientations are along the $c$ axis (easy axis) and the magnetic moments in one channel are aligned antiparallel each other. Moreover, between two channels, the magnetic moments on the inversion-related Mn atoms (connected by dotted line in Fig. 1(a)) are also aligned along the opposite directions\cite{TangP}. Thus, both $P$ and $T$ symmetries are broken but the combined $PT$ symmetry is preserved. The result of single crystal XRD indicates that the obtained crystals have an orthorhombic symmetry and the determined lattice parameters are $a=$ 6.587(6) \AA, $b=$ 3.870(5) \AA, and $c=$ 7.315(8) \AA, consistent with previous results of $o$-CuMnAs polycrystal\cite{Mundelein,Maca}.
The plate-like crystal with rectangular shape (inset of Fig. 1(b)) is consistent with the orthorhombic symmetry of $o$-CuMnAs. As shown in Fig. 1(b), all of peaks in the XRD pattern of a single crystal can be indexed by the indices of $(l00)$ lattice planes of $o$-CuMnAs. It reveals that the crystal surface is normal to the $a$-axis with the plate-shaped surface parallel to the $bc$ plane. Furthermore, the single crystal XRD shows that the long edge of the plate-like crystal is parallel to the crystallographic $b$ axis. The EDX spectrum of a single crystal confirms the presence of Cu, Mn, and As and the absence of Bi. The average atomic ratio determined from EDX is Cu : Mn : As = 0.96(1) : 0.97(1) : 1.000(6) when setting the content of As as 1. It indicates that the Cu and Mn positions are almost fully occupied and there are only very small amount of vacancies in both sites.

\begin{figure}[tbp]
\centerline{\includegraphics[scale=0.43]{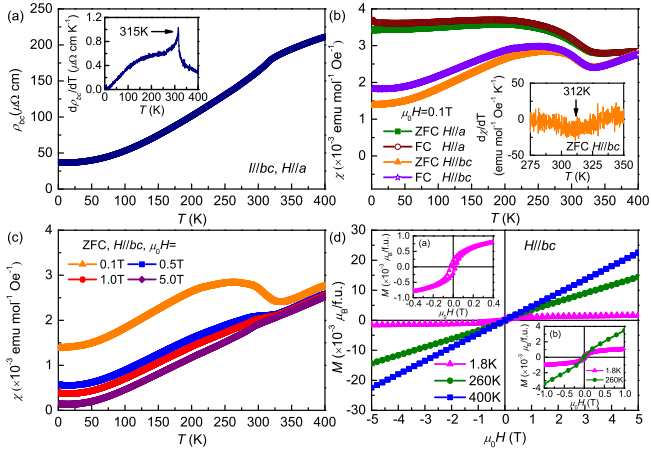}} \vspace*{-0.3cm}
\caption{(a) Temperature dependence of resistivity $\rho_{bc}(T)$ for $o$-CuMnAs single crystal with $I\Vert bc$. Inset: $d\rho_{bc}(T)/dT$ as a function of $T$. (b) Temperature dependence of magnetic susceptibility $\chi(T)$ with zero-field-cooling (ZFC) and field-cooling (FC) modes at $\mu_{0}H=$ 0.1 T for $H\Vert a$ and $H\Vert bc$. Inset: $d\chi(T)/dT$ vs. $T$ with ZFC mode for $H\Vert bc$. (c) ZFC $\chi(T)$ curves at various fields for $H\Vert bc$. (d) Isothermal $M(H)$ curves at 1.8 K, 260 K and 400 K for $H\Vert bc$. Inset: (a) enlarged part at $|\mu_{0}H|\leq$ 0.4 T at $T=$ 1.8 K. (b) enlarged part at $|\mu_{0}H|\leq$ 1.0 T at $T=$ 1.8 K and 260 K.}
\end{figure}

Temperature dependence of resistivity $\rho_{bc}(T)$ in the $bc$-plane for $o$-CuMnAs single crystal is shown in Fig. 2(a). It exhibits a metallic behavior in the whole temperature range of measurement. Moveover, there is a change of slope at $T\sim$ 315 K determined from the maximum position of $d\rho_{xx}(T)/dT$ curve. This is corresponding to the AFM transition temperature ($T_{N}$) shown below. It has to be noted that although the temperature where the resistivity kink appears is consistent with that in $o$-CuMnAs polycrystal\cite{Maca}, the behavior of resistivity above $T_{N}$ are slightly different: the metallic behavior persists up to 400 K for single crystal when the weakly temperature dependent behavior with negative slope appears for polycrystal. Such difference could be related to the effects of grain boundary or the anisotropy of resistivity. Much larger resistivity of polycrystal\cite{Maca} than that in single crystal could partially reflect these effects.

Fig. 2(b) exhibits magnetic susceptibility $\chi(T)$ with zero-field-cooling (ZFC) and field-cooling (FC) modes at $\mu_{0}H=$ 0.1 T for $H\Vert a$ and $H\Vert bc$.
For both field directions, the $\chi(T)$ curves show obvious kinks at $T\sim$ 312 K determined from the peak of derivative $d\chi(T)/dT$ curve (inset of Fig. 2(b)). This anomaly is corresponding to the $T_{N}$ and in good agreement with the results of $\chi(T)$ in polycrystal\cite{Maca} and resistivity (heat capacity) shown above (below). Secondly, the $\chi(T)$ for $H\Vert bc$ decreases with lowering temperature, which is typical behavior for antiferromagnet. In contrast, the $\chi(T)$ for $H\Vert a$ is almost temperature independent below $T_{N}$. Moreover, the former has smaller absolute values than the latter. It suggests that the spin direction at $T<T_{N}$ is in the $bc$ plane and ordered antiferromagnetically each other. The upturn transition at $\mu_{0}H=$ 0.1 T becomes weaker when increasing field and the shapes of $\chi(T)$ curves at high fields are similar to that in polycrystal\cite{Maca}. Thirdly, it can be seen that there is no Curie-Weiss behavior at $T>T_{N}$ for both field directions, suggesting that there may be an AFM fluctuation at high temperature region as observed in polycrystal\cite{Maca}. Finally, the ZFC and FC $\chi(T)$ curves overlap each other rather well, indicating the absence of spin-glass state in $o$-CuMnAs single crystal. Fig. 4(c) plots the field dependence of magnetization $M(\mu_{0}H)$ at $T=$ 1.8 K, 260 K and 400 K for $H\Vert bc$. At $T>T_{N}$ (400 K), the $M(\mu_{0}H)$ curve shows a paramagnetic behavior, i.e., linear dependence of $M(\mu_{0}H)$ on magnetic field.
In contrast, there is a small but finite hysteresis at low temperatures superimposing on the linear field-dependent signal. At 1.8 K, the coercive field $\mu_{0}H_{c}$ is close to 0.0276 T and the saturated moment $M_{s}$ at 0.4 T is about 8$\times$10$^{-4}$ $\mu_{\rm B}$/f.u. (inset (a) of Fig. 2(d)). Although the $M_{s}$ at 1.8 K is small, but it should not originate from the extrinsic FM impurities because the hysteresis behavior is still observable at 260 K but disappears at 400 K (inset (b) of Fig. 2(d)), in agreement with the magnetic transition temperature. Both the upturn of $\chi(T)$ curve at low field (Fig. 2(b)) and the hysteresis of $M(H)$ curve below $T_{N}$ suggests that the long-range AFM order might be canted, leading to the weak FM behavior.

\begin{figure}[tbp]
\centerline{\includegraphics[scale=0.43]{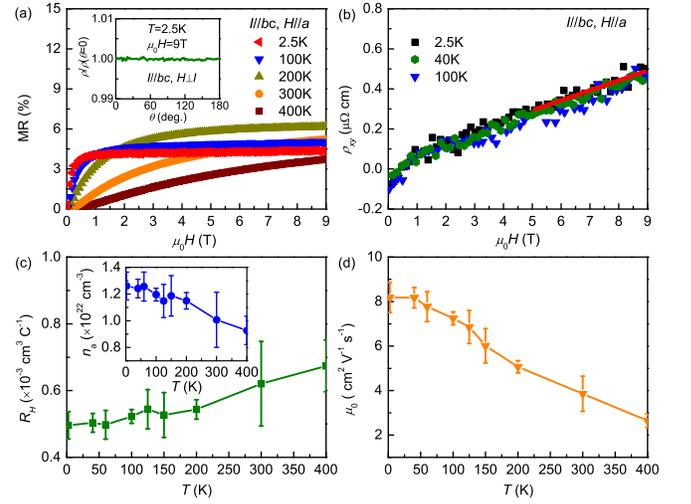}} \vspace*{-0.3cm}
\caption{Field dependence of (a) magnetoresistance (MR) and (b) Hall resistivity $\rho_{xy}(\mu_{0}H)$ up to $\mu_{0}H$ = 9 T for $H\parallel a$ at various temperatures. Inset of (a): angular-resolved MR at $\mu_{0}H$ = 9 T and $T=$ 2.5 K. The red solid line in (b) represents the linear fit of $\rho_{xy}(\mu_{0}H)$ curve between 5 - 9 T at 2.5 K. (c) Temperature dependence of Hall coefficient $R_{H}(T)$ at $\mu_{0}H$ = 9 T. Inset: temperature dependence of apparent carrier concentration $n_{a}(T)$ obtained from the single-band model. (d) Derived carrier mobility $\mu_{0}(T)$ as a function of temperature.}
\end{figure}

Fig. 3(a) shows the MR ($=(\rho_{xx}(T,\mu_{0}H)-\rho_{xx}(T,0))/\rho_{xx}(T,0)$) of $o$-CuMnAs single crystal at various temperatures. At 2.5 K, the MR increases with field quickly and then saturate at high field. According to the two-band model, it suggests that one type of carriers has a much higher concentration than another type\cite{Pippard}. With increasing temperature, the saturated value of MR is similar to that at low temperature but the saturation field becomes larger, which can be partially ascribed to the decrease of carrier mobilities at high temperature\cite{Pippard}. The MR of $o$-CuMnAs single crystal is very small ($\sim$ 4 - 6 \% at 9 T), much smaller than those in most of known TSMs, such as Cd$_{3}$As$_{2}$, TaAs, and WTe$_{2}$\cite{LiangT,HuangXC,Ali}. Moreover, the saturation behavior of MR in $o$-CuMnAs is also remarkably different from those in reported TSMs, which usually exhibit unsaturated MR even at very high field. Inset of Fig. 3(a) shows the angular-resolved MR (ARMR) at $\mu_{0}H$ = 9 T and $T=$ 2.5 K. The $\theta$ is defined as the angle between magnetic field and the $a$ axis and the $\theta=$ 90$^{\circ}$ is corresponding to $H\Vert bc$. The field is always perpendicular to the current direction. The ARMR is insensitive to the field direction, suggesting that the scattering time and/or effective mass of carriers $m^{*}$ are almost isotropic.

In order to get more information on the carriers in $o$-CuMnAs single crystal, the field dependence of Hall resistivity $\rho_{xy}(\mu_{0}H)$ at various temperatures is investigated. As shown in Fig. 3(b), the $\rho_{xy}(\mu_{0}H)$ shows a linear dependence on magnetic field up to 9 T and it barely depends on the variation of temperatures. The positive slopes of $\rho_{xy}(\mu_{0}H)$ curves indicate that the dominant carriers are hole-type in the whole temperature range of measurement. The corresponding Hall coefficient $R_{H}(T)\equiv\rho_{xy}(\mu_{0}H)/\mu_{0}H$ determined from the linear fits of $\rho_{xy}(T,\mu_{0}H)$ curves between 5 - 9 T also shows a relatively weak temperature dependence (Fig. 3(c)).
Based on the single-band model $R_{H}=1/|e|n_{a}$, the apparent carrier concentration $n_{a}$ can be estimated (inset of Fig. 3(c)). The $n_{a}$ at 2.5 K is about 1.2(1) $\times$ 10$^{22}$ cm$^{-3}$ and decreases slightly with increasing temperature. It should be noted that the $n_{a}$ in $o$-CuMnAs is much larger than those in many of TSMs, such as Cd$_{3}$As$_{2}$, TaAs, and NbP\cite{LiangT,HuangXC,Shekhar}. It is even larger than those of nodal-line semimetals, such as ZrSiSe and ZrSiTe\cite{HuJ}. Small isotropic MR with saturation behavior at high field, linear field dependence of $\rho_{xy}(\mu_{0}H)$ with positive slope, and weak temperature dependence of $n_{a}$ strongly suggest that $o$-CuMnAs behaves like a single-band metal with dominant hole-type carriers. According to the single-band model, the carrier mobility $\mu_{0}$ can be derived from the formula $\mu_{0}=\sigma_{xx}(0)/|e|n_{a}\approx 1/|e|n_{a}\rho_{xx}(0)=R_{H}/\rho_{xx}(0)$. The derived $\mu_{0}$ as a function of temperature is shown in Fig. 3(d). The $\mu_{0}(T)$ increases monotonically with decreasing temperature. This behavior is often observed when the $e-ph$ scattering is dominant. Because the phonon is frozen gradually at low temperature, the $e-ph$ scattering rate decreases, leading to the increase of carrier mobility\cite{Ziman}. Although the $\mu_{0}$ increases with lowering temperature, it is only 8.1(7) cm$^{2}$ V$^{-1}$ s$^{-1}$ at 2.5 K, which is about 3 - 6 orders of magnitude smaller than those in other TSMs, such as Cd$_{3}$As$_{2}$, TaAs, NbP, and WTe$_{2}$\cite{LiangT,HuangXC,Shekhar,LuoY}.

\begin{figure}[tbp]
\centerline{\includegraphics[scale=0.27]{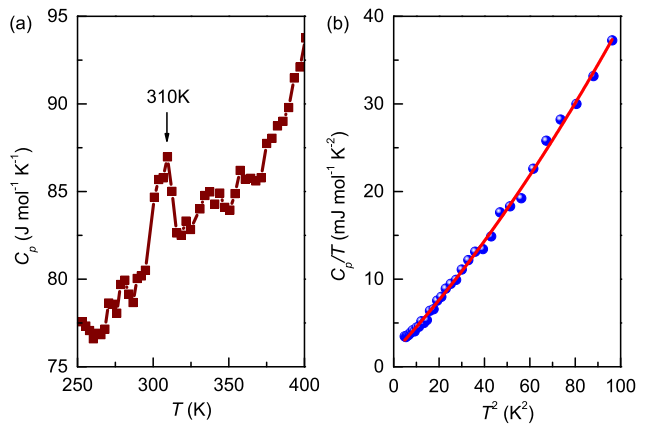}} \vspace*{-0.3cm}
\caption{(a) Temperature dependence of $C_{p}(T)$ for $o$-CuMnAs single crystal at high temperature region (250 - 400 K). (b) The relationship between $C_{p}/T$ and $T^{2}$ at low temperature region. The red solid curve represents the fitting result using the formula $C_{p}/T=\gamma+\beta T^{2}+\delta T^{4}$.}
\end{figure}

Fig. 4(a) shows the high-temperature heat capacity $C_{p}$ of $o$-CuMnAs single crystal as a function of temperature. There is a peak with the maximum value located at 310 K, consistent with the $T_{N}$ obtained from magnetization and resistivity measurements. It clearly indicates that the AFM transition in $o$-CuMnAs at $T_{N}\sim$ 312 K is bulk and long-ranged. Fig. 4(b) exhibits the relationship between $C_{p}/T$ and $T^{2}$ at low temperature region. In order to extract the electronic specific heat coefficient $\gamma$ and Debye temperature $\Theta _{D}$, the curve is fitted using the formula $C_{p}/T=\gamma+\beta T^{2}+\delta T^{4}$. The fit is rather good with $\gamma$ = 1.6(2) mJ mol$^{-1}$ K$^{-2}$, $\beta$ = 0.28(1) mJ mol$^{-1}$ K$^{-4}$, and $\delta$ = 9(1)$\times$10$^{-4}$ mJ mol$^{-1}$ K$^{-6}$. When using the free electron model and obtained values of $\gamma$ and $n_{a}$ at 2.5 K, the estimated $m^{*}$ is 1.5(2) $m_{e}$, where $m_{e}$ is the free electron mass. The $m^{*}$ of $o$-CuMnAs is much larger than the experimental values of $m^{*}$ in known TSMs (usually in the range of 0.05 - 0.1 $m_{e}$)\cite{Shekhar,GuoST,XiongJ2}. On the other hand, the $\Theta _{D}$ is estimated to be 274(4) K according to the formula $\Theta _{D}$ = $(12\pi^{4}NR/5\beta )^{1/3}$, where $N$ is the atomic number in the chemical formula ($N$ = 3) and $R$ is the gas constant ($R$ = 8.314 J mol$^{-1} $ K$^{-1}$).
It has to be mentioned that because the AFM spin wave excitation (magnons) might have some contributions to the heat capacity and it follows a $T^{3}$ behavior at low temperature, there might be some uncertainty when evaluating the $\Theta _{D}$, but the fitted value of $\gamma$ should still be reliable.

The large $n_{a}$ and $m^{*}$, the low mobility and small MR imply that the dominant carriers in $o$-CuMnAs might not behave like relativistic fermions. The contradiction between experimental results and theoretical calculations could be due to the following possible reasons. The first one is the gapping of Dirac cone near $E_{F}$ due to the rotation of spin orientation away from the $c$ axis. As pointed in the theoretical calculation, if the spin orientation is along (111) or (101) direction, both $R_{y}$ and $S_{2z}$ symmetries will be broken and a small gap will be opened\cite{TangP,Smejkal}. The canted AFM observed in experiment may have similar influence on the band structure, leading to the opening of gap near Dirac points. The second one is the electronic correlation effect which could also gap the Dirac cones and introduce the finite mass to carriers\cite{TangP}. The third one is the position of $E_{F}$. Because of small amount of disorders or deficiencies in Cu and/or Mn sites, the $E_{F}$ could move far away from the crossing point of Dirac cone and the carriers may not be regarded as massless Dirac fermions but behave like massive fermions. The angle-resolved photoemission spectroscopy measurement is urgently needed to verify the experimental electronic structure and surface state of $o$-CuMnAs in the future.

\section{Conclusion}

In summary, we have successfully grown $o$-CuMnAs single crystals using the Bi flux method. $o$-CuMnAs shows a metallic behavior with an AFM transition above room temperature ($T_{N}\sim$ 312 K). The measurements of magnetic properties suggest that this long-range AFM order could be canted with spin orientation in the $bc$ plane. On the other hand, $o$-CuMnAs exhibits a very small and nearly isotropic MR even at low temperature. Hall measurements indicate that the hole-type carriers are dominant with rather high concentration. Moreover, the carriers in $o$-CuMnAs have a large effective mass and much low mobility when compared to those in typical TSMs. These results suggest that the carriers in $o$-CuMnAs behave more like nonrelativistic massive fermions not Dirac ones predicted in theory.

\section{Acknowledgments}

This work was supported by the Ministry of Science and Technology of China (2016YFA0301300, 2016YFA0300504), the National Natural Science Foundation of China (Grant No. 11574394, 11774423), and the Fundamental Research Funds for the Central Universities (No. 2017RC20, 2017PTB-03-04), and the Research Funds of Renmin University of China (RUC) (15XNLF06, 15XNLQ07).

Note added.-After submitted our manuscript, we become aware of a preprint studying on the physical properties, especially the magnetic structure of $o$-CuMnAs\cite{Emmanouilidou}. Both work are consistent with each other. The neutron diffraction data in that work indicates the spin direction is along the $b$ axis, leading to the formation of gap at Dirac point and the gain of mass for electrons in $o$-CuMnAs, in agreement with our main conclusion derived from the analysis of combined results of transport and heat capacity measurements.

\end{document}